\documentclass[aps,prl,groupedaddress,reprint,showpacs,pdftex]{revtex4-1}
\usepackage{graphicx}
\usepackage{tabularx}
\usepackage{bm}
\usepackage{amsmath}
\usepackage{amssymb}
\usepackage{txfonts}
\usepackage{url}
\usepackage[usenames,dvipsnames]{color}
\definecolor{Gray}{gray}{0.0}
\definecolor{lightGray}{gray}{0.35}

\begin{document}
\title{
  {\it Ab initio} evaluation of complexation energies
  for cyclodextrin drug inclusion complexes
}

\author{Kenji Oqmhula$^{1}$}
\email{mwkokk1907@icloud.com}
\author{Kenta Hongo$^{2, 3, 4, 5}$}
\author{Ryo Maezono$^{5, 6}$}
\author{Tom Ichibha$^{6}$}

\affiliation{$^{1}$
  School of Materials Science, JAIST, 1-1 Asahidai, Nomi, Ishikawa, 923-1292, Japan
}

\affiliation{$^{2}$
  Research Center for Advanced Computing Infrastructure, JAIST, 1-1 Asahidai, Nomi, Ishikawa 923-1292, Japan
}
\affiliation{$^{3}$
  Center for Materials Research by Information Integration, Research and Services Division of Materials Data and Integrated System, National Institute for Materials Science, 1-2-1 Sengen, Tsukuba 305-0047, Japan
}
\affiliation{$^{4}$
  PRESTO, Japan Science and Technology Agency, 4-1-8 Honcho, Kawaguchi-shi, Saitama 322-0012, Japan
}
\affiliation{$^{5}$
  Computational Engineering Applications Unit, RIKEN, 2-1 Hirosawa, Wako, Saitama 351-0198, Japan
}

\affiliation{$^{6}$
  School of Information Science, JAIST, 1-1 Asahidai, Nomi, Ishikawa, 923-1292, Japan
}
\affiliation{$^{7}$
  Computational Engineering Applications Unit, RIKEN, 2-1 Hirosawa, Wako, Saitama 351-0198, Japan
}
\affiliation{$^{8}$
  School of Information Science, JAIST, 1-1 Asahidai, Nomi, Ishikawa, 923-1292, Japan.
}

\begin{abstract}
  We examined the reliability of exchange-correlation functionals 
  for molecular encapsulations combined by van der Waals forces, 
  comparing their predictions with those of diffusion Monte Carlo method. 
  We established that functionals with D3 dispersion force correction
  and including sufficient proportion of exact-exchange in long-ranged
  interaction can comparatively reliably estimate the binding strength.
  Our finding agrees with a previous \textit{ab initio} study on argon dimer. 
  However we found that even such functionals may not be able to
  distinguish the energy differences among different conformations.
\end{abstract}
\maketitle
\newpage

\section{Introduction}
\label{sec.introduction}
Biopharmaceuticals are manufactured, extracted, and semi-synthesized
from biological sources. They have high potency at low dose
and distinctive medical properties in general,~\cite{2014MIT, 2006GEO}
compared to conventional chemical pharmaceuticals.
On the other hand, they often lack physical and/or chemical stabilities,
making some difficulties, for example, they cannot be stored stably
for a long period and cannot be administered orally.
To such problems, one of the most promising solutions is
molecular encapsulation:~\cite{2006GEO,2014MIT} 
The biopharmaceutical molecule (guest) is combined with the carrier
molecule (host), and it gets stabilized physically and chemically. 
It also makes <it possible to control the absorption location and timing.
~\cite{2008VYA} 

\vspace{2mm} 
These properties heavily depend on the binding strength 
between the guest and host molecules. 
Therefore, if one can reliably predict the binding energy 
from simulation, 
it could significantly accelerate the development of 
the molecular encapsulation technique. 
Although the most promising simulation tool on this purpose
is density functional theory (DFT), it is still difficult
to describe the molecular encapsulations using exchange-correlation
(XC) functionals, because the binding consists of various non-covalent
forces: hydrogen bonding, dispersion force, hydrophobic interaction,
and so on.~\cite{2016YE} Several special ideas to describe the non-covalent
interactions have been suggested so far. A promising way is long-range
correction,~\cite{2001IIK,2004YAN} which enhances a proportion of
the exact-exchange in long-range interactions and improves the description
of van der Waals forces.~\cite{2012TSU}
An other choice is using Minnesota functionals, whose parameters
are trained for both covalent and non-covalent systems unlike B3LYP.
~\cite{2007ZHA}
While this idea significantly enhances the reliability for non-covalent
systems, these functionals have an apparent defect that it cannot reproduce
the asymptotic decrease of van der Waals forces.~\cite{2017HON}
The most popular way to describe the damping would be Grimme's dispersion
force corrections (D3).~\cite{2010GRI}, which employs an empirical function
akin to the Lennard-Jones potentials.~\cite{2011GRI}

\vspace{2mm}
In this paper, we examine the reliability of the functionals 
listed in Table~\ref{tab.functionals} to evaluate the binding energy 
between cyclodextrins (host) and plumbagin (guest).
The cyclodextrins (CDs) are ones of the most important
host molecules in molecular encapsulation techniques 
due to its various advantages (see section \ref{sec.system}).
However it has not been studied which functionals can reliably
describe the encapsulation process by the CDs. 
We compared the predictions by each XC functional
with those by diffusion Monte Carlo (DMC) method,
to evaluate their performances.

\vspace{2mm}
The following part of this paper is composed as follows. 
In section \ref{sec.system}, we will introduce about
our target systems, CDs and plumbagin.
In section \ref{sec.details}, we will explain how we obtained
the binding geometries using a docking analysis.
Also we explain the details of our DFT and DMC calculations. 
In section \ref{sec.results}, we show our results and make discussions.
Finally we summarize this paper in section \ref{sec.conc}.

\begin{table*}[htbp]
  \begin{center}
    \caption{\label{tab.functionals}
      List of XC functionals we tested. 
      We examined the reliability of generally used functionals,
      B3LYP, M06L, and M06-2X, and their relatives with
      D3~\cite{gd3} and/or CAM~\cite{cam} corrections,
      for predicting the binding energies between
      plumbagin and cyclodextrins.
    }
    \begin{tabular}{c|c|c|c}
      Plain  & D3         & CAM       & CAM+D3       \\
      \hline
      B3LYP  & CAM-B3LYP  & B3LYP-D3  & CAM-B3LYP-D3 \\
      M06L   & M06L-D3    & --        & --            \\
      M06-2X & M06-2X-D3  & --        & --            \\
    \end{tabular}
  \end{center}
\end{table*}
\section{System}
\label{sec.system}
Plumbagin is an organic molecule including two benzene rings.~\cite{1999OOM} 
This molecule is known to have a medical efficacy against prostate cancers,
~\cite{2008AZI,2017ABE} but the difficulty of its storage hinders
its spread at practical level: 
63.8\% of plumbagin is lost in one month under atmospheric condition
due to oxidation and degradation.~\cite{2010SUT}
For the problem, the molecular encapsulation by the CDs could be
the most promising solution.~\cite{2010SUT}

\begin{figure}[htbp]
  \centering
  \includegraphics[width=1.0\hsize]{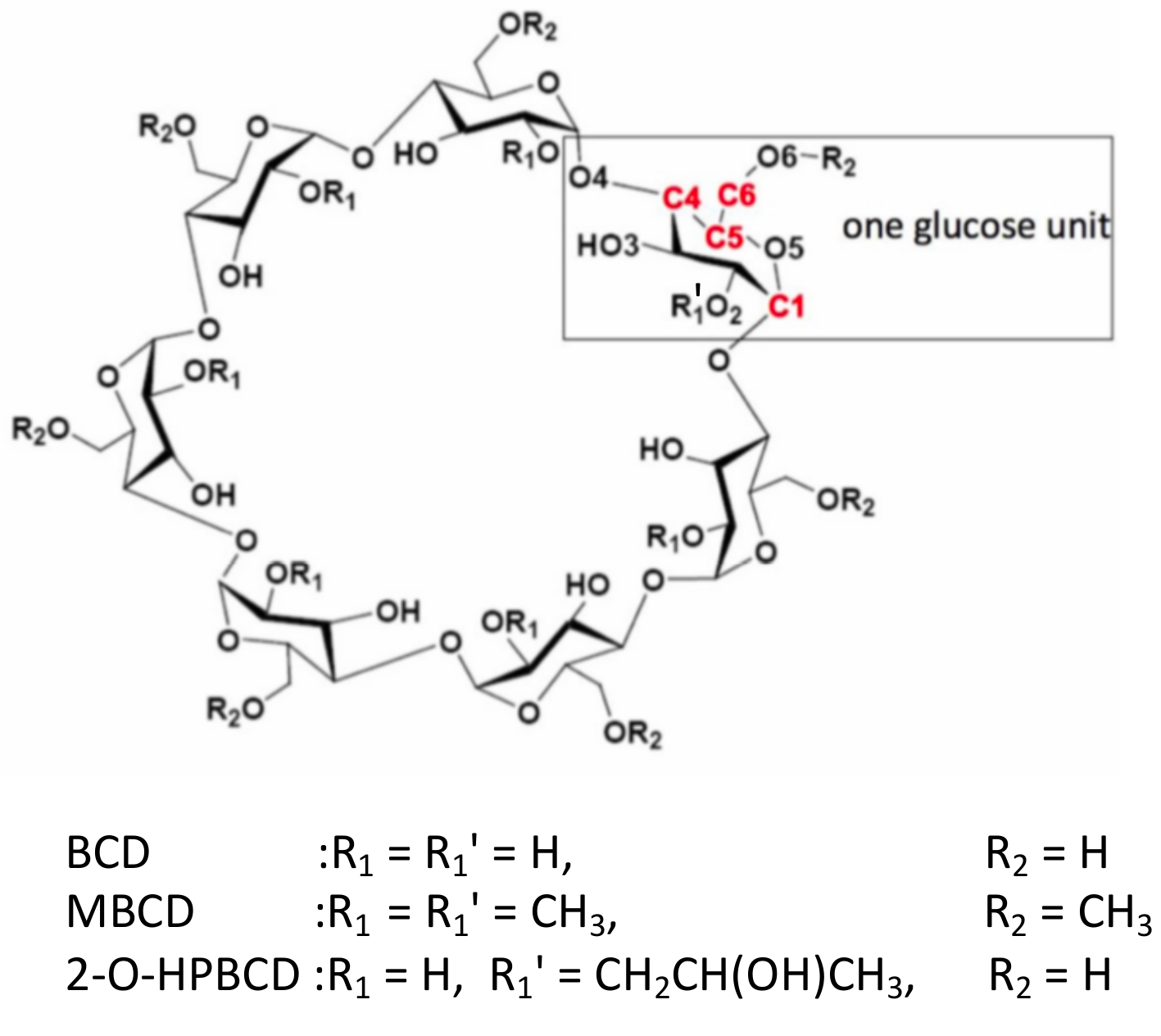}
  \caption{
    \label{fig.structure}
    The molecular structure of BCDs. The ring consists of seven
    glucose units. There are various BCDs with different functionals
    groups, and we selected BCD, MBCD, and 2-O-HPBCD shown in this
    figure. Here, in the case of 2-O-HPBCD, just one CH$_2$CH(OH)CH$_3$
    puts on one of the $R_1$, and the other $R_1$ are all hydrogen atoms.  
  }
\end{figure}

\vspace{2mm}
CD is a circular molecule consisting of glucose units as shown
in Figure~\ref{fig.structure} and is broadly used as a carrier
of various pharmaceuticals due to the following benefits other
than stabilization:~\cite{2008VYA}
\begin{itemize}
\item
  The circle size is adjustable to the size of guest molecule
  by changing the number of glucose units $n$ ($\ge$6).
  When $n$ is 6, 7 and 8, CD is called $\alpha$-, $\beta$-,
  and $\gamma$-CD, respectively.
\item
  Docking with CD also improves the drug solubility or dissolution,
  which helps the adsorption of the drugs.
\item
  The release rate/timing is controllable
  by replacing the functional groups. 
\end{itemize}
The suitable ring size for plumbagin is $\beta$-CD (BCD),~\cite{1999OOM}
and we calculated the binding energies between plumbagin and
some representative CDs, BCD, Methyl-BCD (MBCD), and 2-O-HPBCD.

\section{Methods}
\label{sec.details}
We obtained the binding structures between plumbagin and BCDs
using docking analysis with Lamarckian algorithm implemented
in AutoDock 4.2.6.~\cite{1998MOR}
This is often used to predict ligand arrangements of protein systems.
A set of genes represents the ligand arrangements and
they are updated to get energetically stable structures.
Here, each of the genes represents translations, orientations,
and conformations of the ligands.
We regarded the plumbagin as the ``ligand'' of BCDs
to run the docking analysis.

\vspace{2mm}
The molecular structures of plumbagin and BCDs are taken
from the entries (PVVAQS01/BCDEXD03/BOYFOK04/KOYYUS) 
in the Cambridge Structural Database.~\cite{2016GRO}
We optimized them using CAM-B3LYP-D3 prior to the docking analysis.
In the analysis, the translation of plumbagin is discretized
on a 50$\times$38$\times$24 grid with a spacing of 0.375~$\AA$.
We updated $\sim$150 genes for 100 iterations.
We selected only one gene at the end of each iteration,
to ``survive'' to the next iteration.
The energy corresponding to each gene was calculated
by an empirical force field, whose electrostatic iteraction is
calculated based on the Gasteringer charges.~\cite{1978GAS}
The other input parameters were set to be the default values
in Autodock 4.2.6. 

\vspace{2mm}
We used Gaussian09/16~\cite{g09, g16} for DFT calculations. 
We performed all-electron calculations with 6-31++G($d$,$p$) 
Gaussian basissets, since the family of 6-31G basis sets are
often used for host-guest docking systems similar to our system.
~\cite{2013BAC, 2014BAC, 2016YE, 2017BOU, 2017DEK}
We corrected the basis set superposition error with
the counterpoise method.~\cite{1996SIM}
We used 8 functionals listed in Table~\ref{tab.functionals}
to calculate the barrier energies, where the geometries
of the isolated plumbagin and BCDs and their compound
are optimized for each of the functionals.


\vspace{2mm}
We applied DMC to evaluate the energies of before/after 
the docking and the binding energy using QMCPACK.~\cite{2018KIM}
These geometries are optimized by DFT calculation with
M06L functional.
Orbital functions used in the Slater determinant are
generated by DFT method with M06L functional implemented in GAMESS.
~\cite{2005DYK,1993SCH}
Core potential in hydrogen atoms were described by
Annaberdiyev's effective core potential~\cite{2018ANN}
and core potential and electrons in carbon and oxygen atoms
were described by Bennett's.~\cite{2017BEN}
We described the Kohn-Sham orbitals with the augmented
cc-pVDZ Gaussian basissets.~\cite{1992KEN}
The Jastrow factor consists of one, two, and three body terms 
amounting to 212 variational parameters in total.
The parameters are optimized by the scheme to minimize
a hybridization of energy and variance with ratio of 7:3
at variational Monte Carlo level~\cite{2001FOU, 2009MAE}.
We estimated a timestep bias by a linear extrapolation
of the energies obtained at two time steps, $dt = 0.020$
and $0.005$ a.u.$^{-1}$.
We set a target population of walkers to be $4,000$.
Practically this size of target population is large enough
to suppress a population control error.

\begin{figure}[htbp]
  \centering
  \includegraphics[width=1.0\hsize]{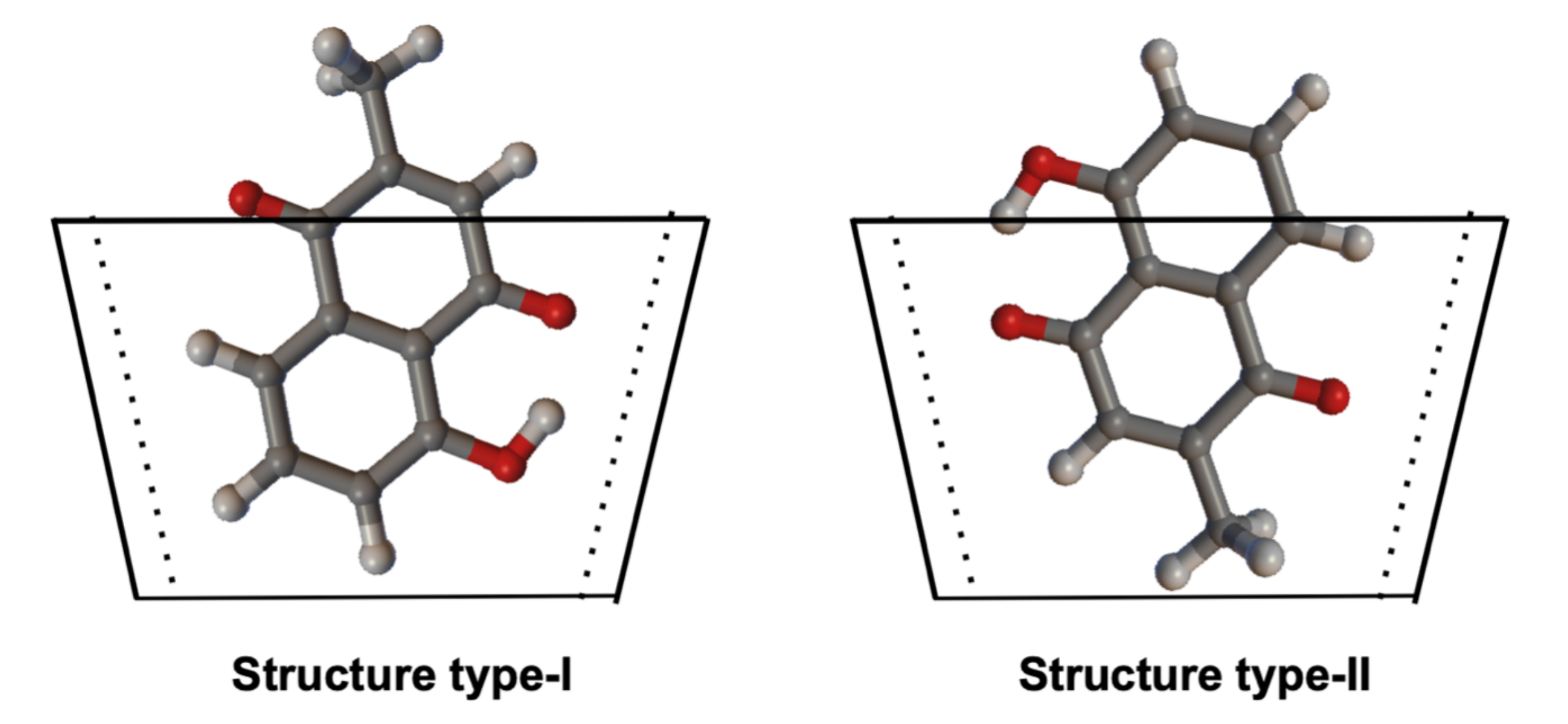}
  \caption{
    \label{fig.up_down}
    Two types of stable conformations found by
    the docking analysis.
    In type-I(II), the hydroxyl phenolic (methyl quinone)
    group of plumbagin is placed around narrow-side of
    the cavity in BCDs.    
  }  
\end{figure}
\begin{figure}[htbp]
  \centering
  \includegraphics[width=1.0\hsize]{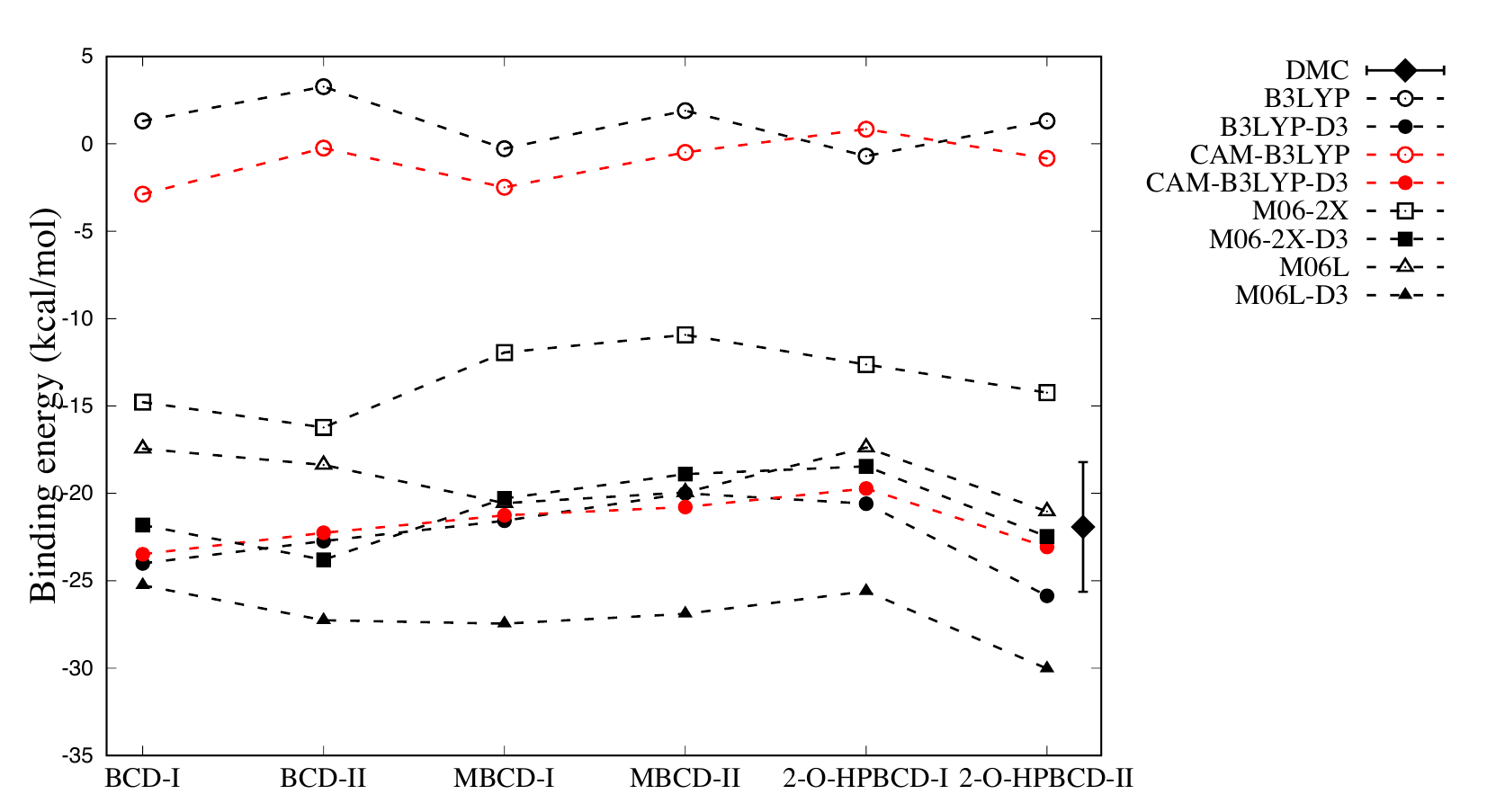}
  \caption{                    
    \label{fig.energies}
    Comparison of the binding energies
    predicted by DMC and DFT with the several functionals.
    DMC was applied only to 2-O-HPBCD-II 
    since the calculation cost of DMC is huge. 
    The difference between type I and type II
    is defined by where the hydroxyl phenolic group
    of the plumbagin is located.
    The detail is described in Figure~\ref{fig.up_down}.
  }
\end{figure}
\section{Results and Discussions}
\label{sec.results}
We found that the structures obtained by the docking analysis
are classified into two types of conformations
as shown in Figure \ref{fig.up_down}.
In type-I~(II) the hydroxyl phenolic (methyl quinone) group of
plumbagin is placed around narrow-side of the cavity in BCDs.
We calculated the binding energies for both conformations.

\vspace{2mm}
Figure \ref{fig.energies} shows the binding energies
evaluated by DFT and DMC.
DMC is applied only to 2-O-HPBCD-II
because of the huge calculation cost.
Firstly, comparing the results of the functionals
without D3 corrections, (CAM-)B3LYP, M06L, and M06-2X,
we can see that M06L and M06-2X does but (CAM-)B3LYP
does not reproduce the stabilization by docking.
The primary reason could be the internal parameters of
(CAM-)B3LYP are trained for only covalent systems.~\cite{2011MAR}
Similar cases are also found for
various types of non-covalent systems.~\cite{2011MAR,2010HON,2012WAT,2013HON,2015HON,2016HON,2018KEN}

\vspace{2mm} 
Secondly, comparing the results of the all functionals we applied, 
the D3 corrected functionals reproduce the DMC result quantitatively,
and M06L-D3 solely predicts significantly larger binding energy 
than DMC. This would be due to the lack of exact-exchange 
in long-ranged interactions. 
This claim is supported by the previous {\it ab initio} study 
that showed insufficient proportion of exact-exchange 
in long-ranged interactions leads an overbind 
in the case of argon dimer.~\cite{2002KAM} 
We can find that B3LYP-D3 also slightly overestimates 
the binding energy, and it can be explained by the same reason, 
because its proportion of the exact-exchange in the long-ranged 
interactions are smaller than CAM-B3LYP-D3 and M06-2X. 
Among the functionals without the D3 correction, 
only M06L quantitatively reproduces the DMC result. 
However this could be just a fortune coming from
that the overbind by the lack of exact-exchange and
the underbind by the lack of dispersion force correction
cancelled out each other.

\vspace{2mm}
Finally we make a discussion focusing on the magnitude
relationships between the binding energies of two conformations,
I and II, predicted by the functionals other than (CAM-)B3LYP
functionals. 
All of the functionals predicted the same magnitude relationships
for MBCD and 2-O-HPBCD.
Of course, it does not directly certify that these predictions
are correct, but generally we can expect they are reliable.
Meanwhile these functionals give different predictions for BCD.
Therefore even D3 corrected functionals with a satisfactory
consideration of long-range exchange interaction cannot
always correctly predict the stablest conformation.
Below we discuss why the contradiction is found just for BCD.

\vspace{2mm}
The contradiction appears between Minnesota and (CAM-)B3LYP-D3 functionals, 
which indicates the contradiction stems from  the difference of the truly 
functional part. Meanwhile it also can be said that the contradiction appears 
among the D3 corrected functionals, which indicates the contradiction comes
from the degree of freedom of the D3 parameters.
In order to clarify which functionals could be correct, 
we examined how significantly the degree of
freedom affects the binding energy predictions in our case.
The D3 parameters can differ depending on the used cost functions in the
parameter fitting to the reference data.
It is reported that the prediction of the binding energy can vary
by utmost 60~kcal/mol depending on the cost function. \cite{2018WEY}
We checked how much our prediction can change depending on
the cost functions in the case of B3LYP-D3, taking a bootstrap analysis
implemented in the BootD3 code. \cite{2018WEY}
As a result, we established that the total energy can be varied
by $\sim$ 10 kcal/mol but the influence on the binding energy is
just less than 0.3~kcal/mol.
Therefore we conclude that the discrepancy of our prediction for BCD
comes from the difference of the truly functional parts. 

\begin{figure}[htbp]
  \centering
  \begin{minipage}{0.45\hsize}
    \centering
    \includegraphics[height=40mm, width=\hsize]{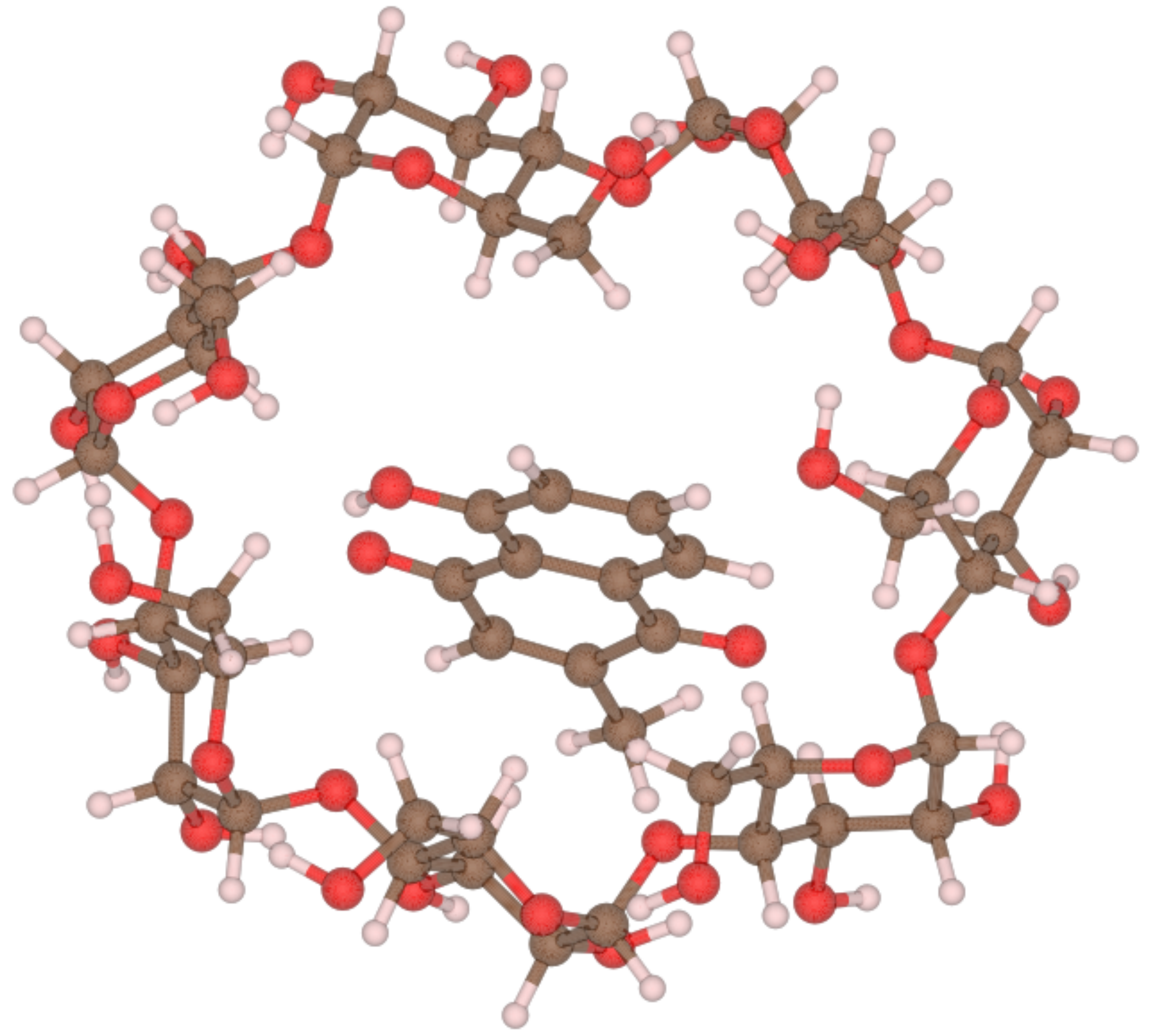}
    BCD-I
  \end{minipage}
  \begin{minipage}{0.45\hsize}
    \centering
    \includegraphics[height=40mm, width=\hsize]{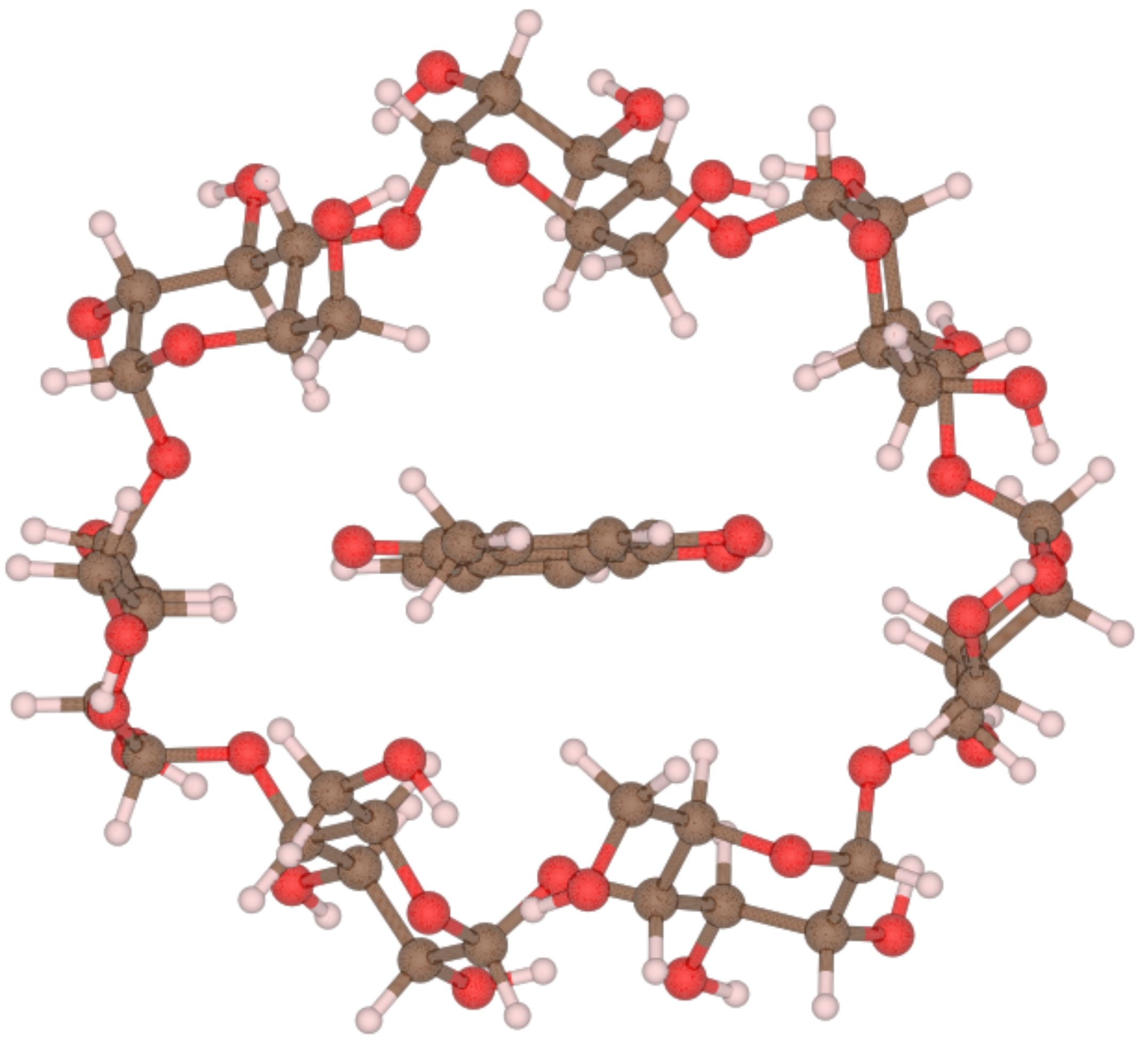} \\
    BCD-II
  \end{minipage}
  \begin{minipage}{0.45\hsize}
    \centering
    \includegraphics[height=40mm, width=\hsize]{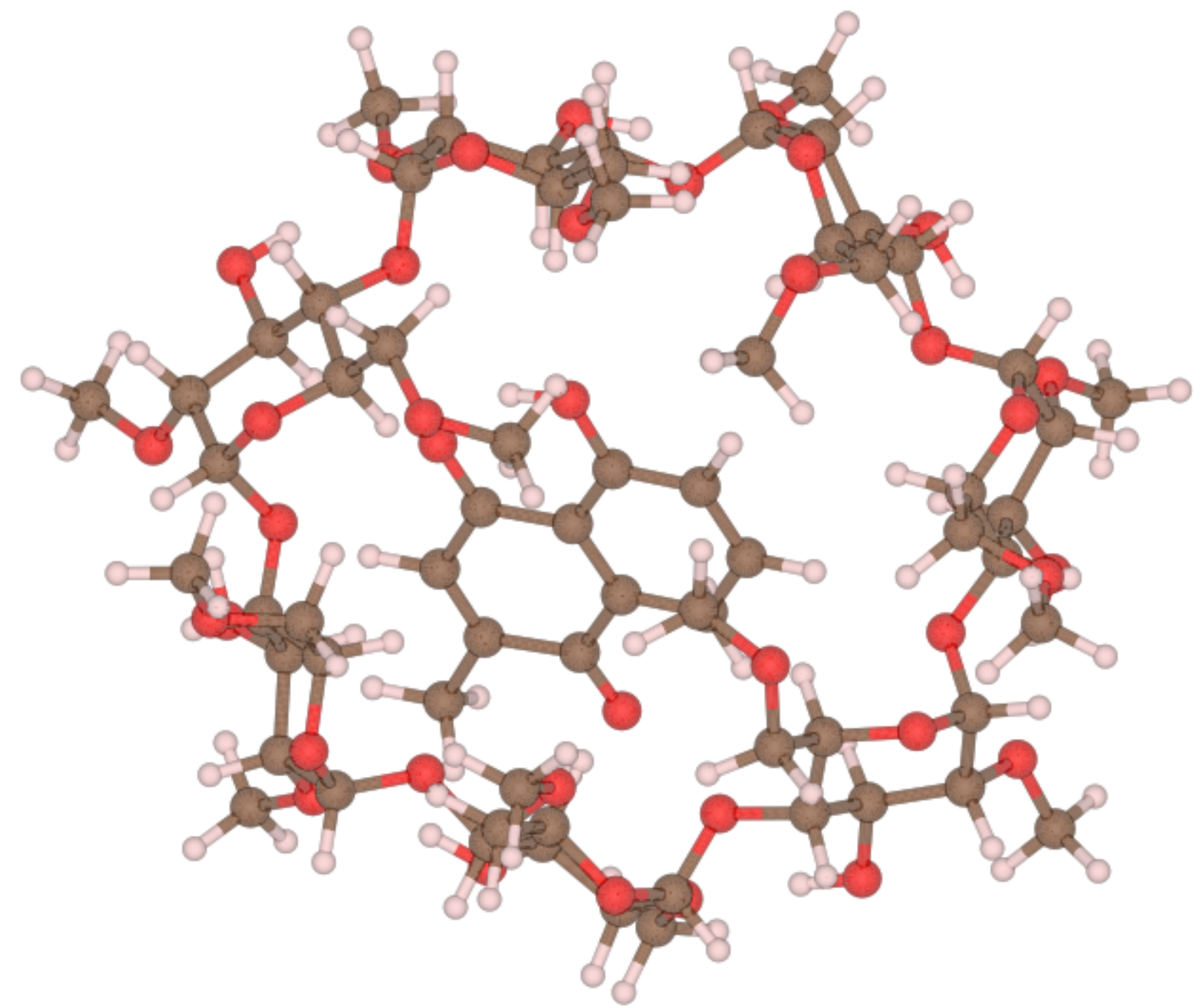}
    MBCD-I
  \end{minipage}
  \begin{minipage}{0.45\hsize}
    \centering
    \includegraphics[height=40mm, width=\hsize]{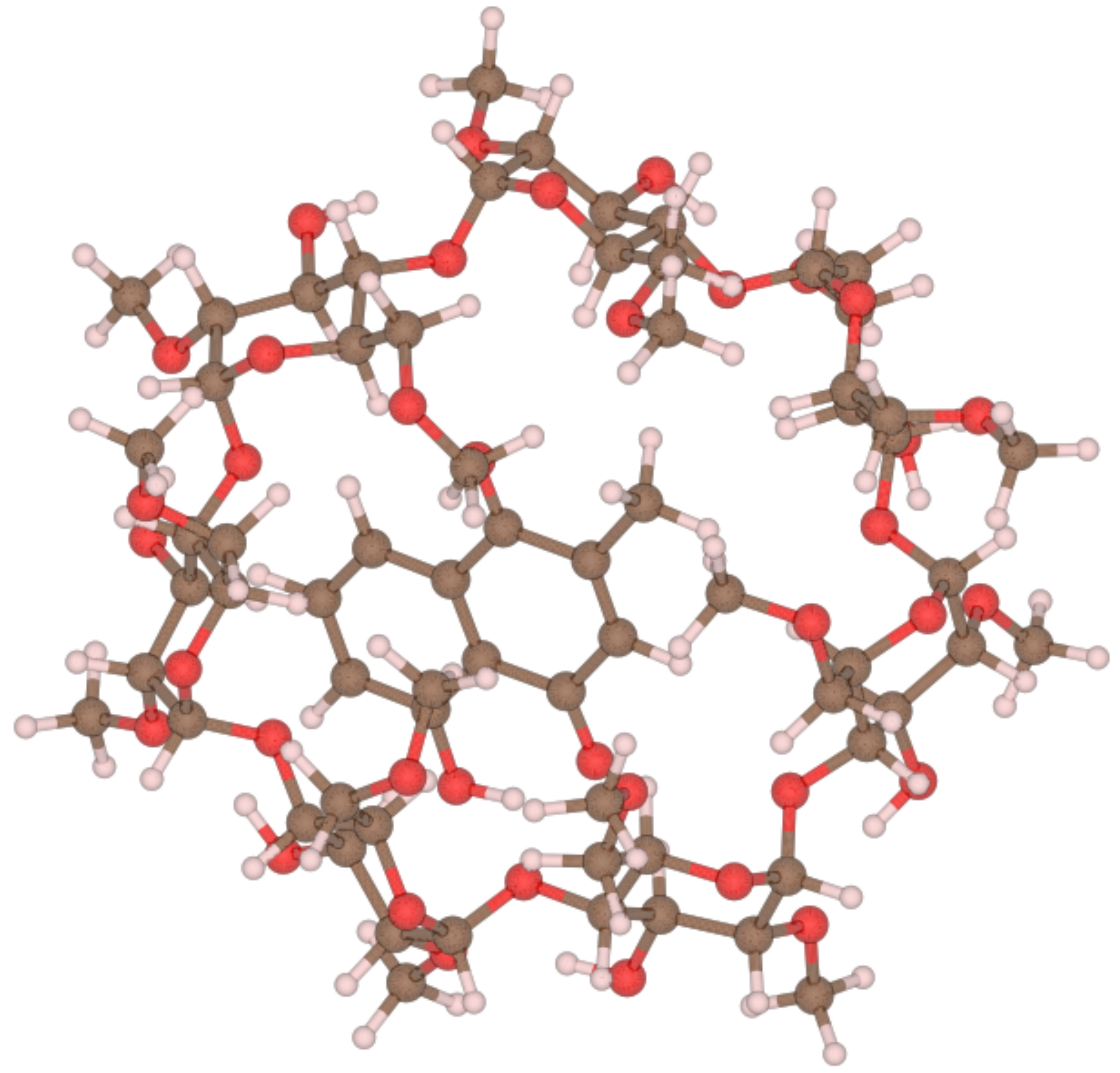} \\
    MBCD-II
  \end{minipage}
  \begin{minipage}{0.45\hsize}
    \centering
    \includegraphics[height=40mm, width=\hsize]{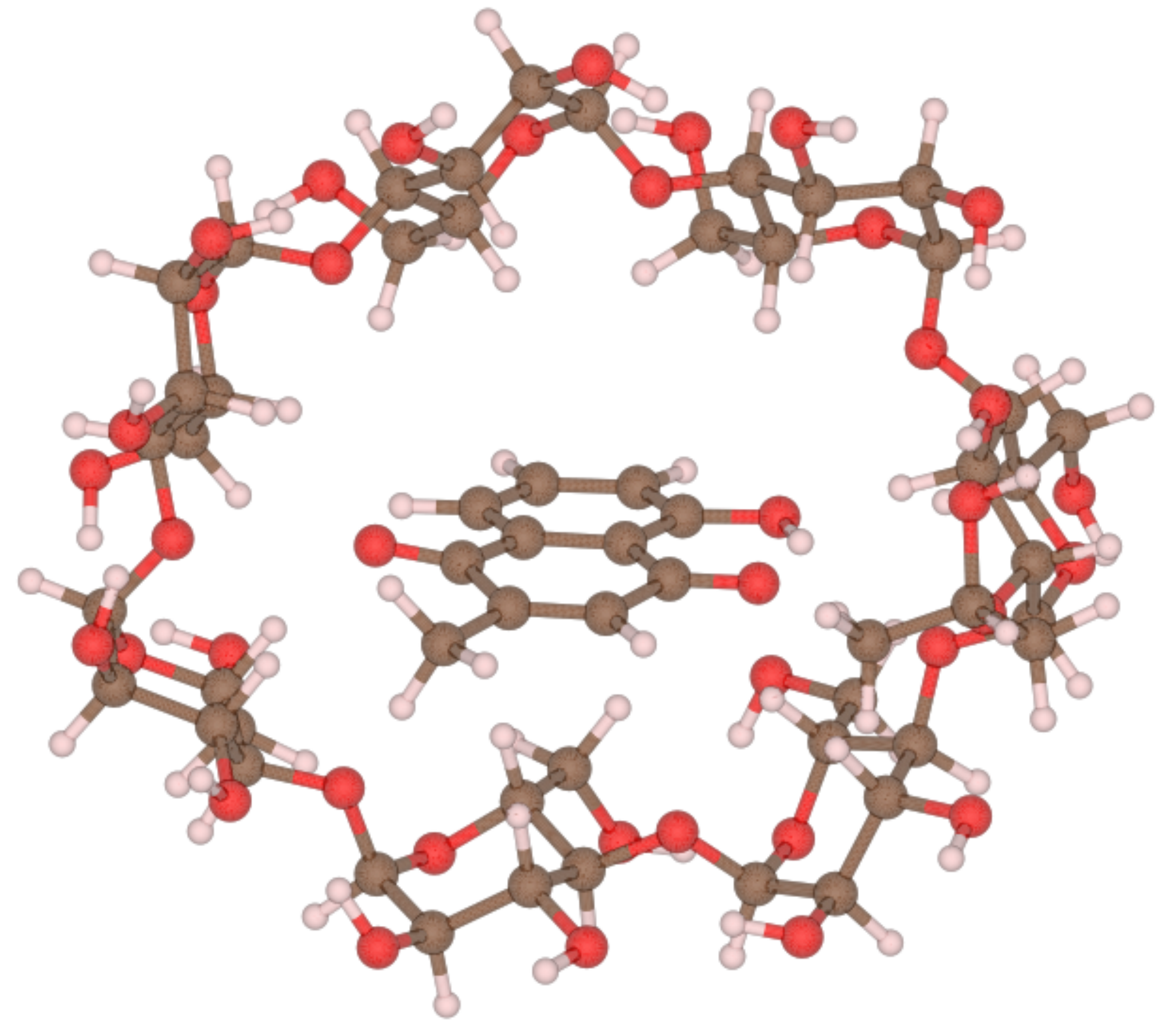}
    2-O-HPBCD-I
  \end{minipage}
  \begin{minipage}{0.45\hsize}
    \centering
    \includegraphics[height=40mm, width=\hsize]{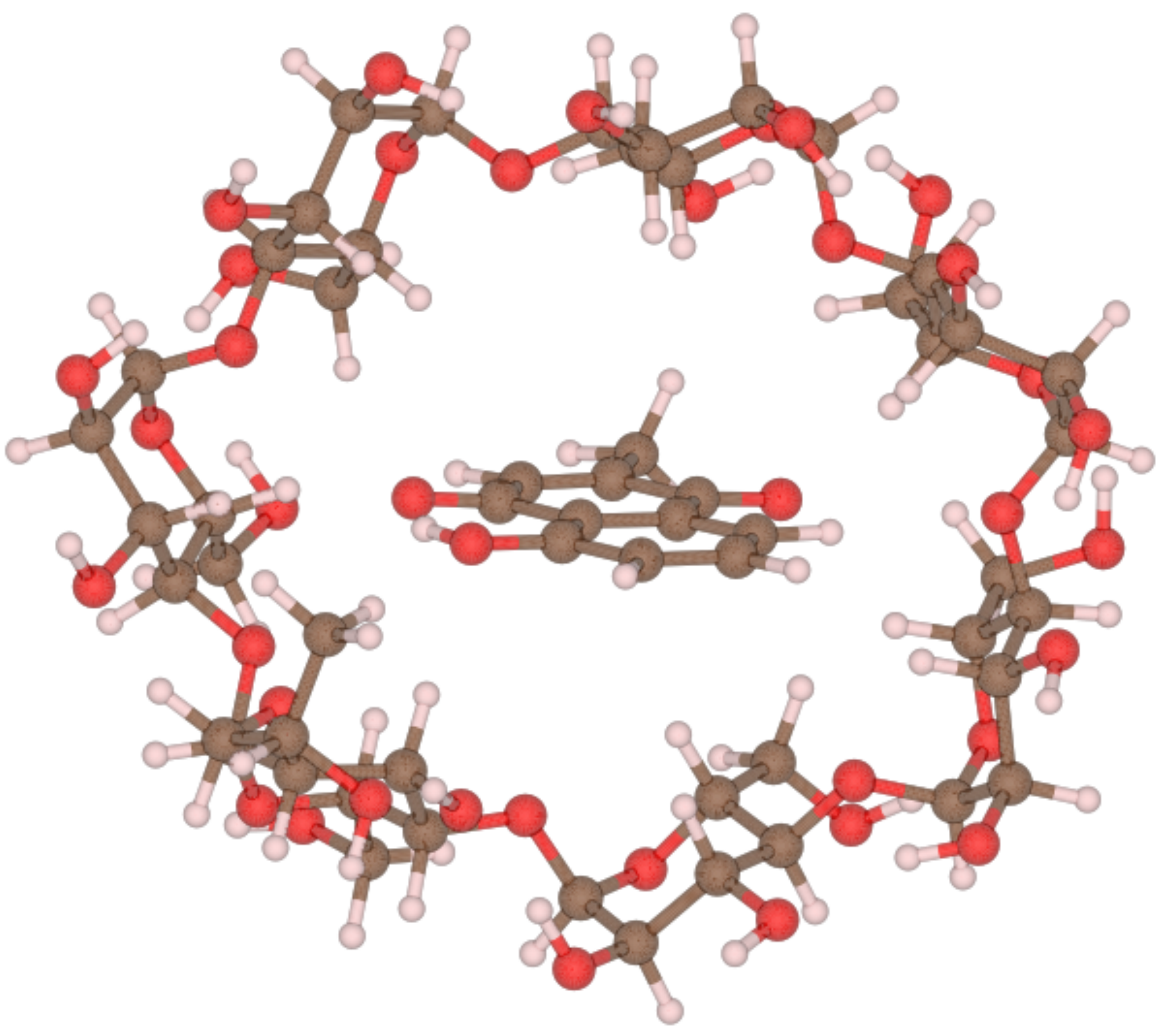} \\
    2-O-HPBCD-II
  \end{minipage}
  \caption{\label{fig.strComp}
    Docking structures between plumbagin and BCDs.
    This figure lists up the docking structures for
    BCD-I, BCD-II, MBCD-I, MBCD-II, 2-O-HPBCD-I, and 2-O-HPBCD-II.
    Those structures are relaxed by CAM-B3LYP-D3.    
  }  
\end{figure}

\vspace{2mm}
Nevertheless we cannot conclude which of Minnesota and (CAM-)B3LYP-D3
functionals is correct only from the DFT results, while we can explain
why the contradiction happens just for BCD.
Figure \ref{fig.strComp} shows the structures of the six conformers.
We found that they can be classified into two patterns of structures:
(a) Plumbagin is in the center of and parallel to the cavity
and (b) plumbagin adheres to the wall of the cavity.
The pattern (b) is especially explained by the CH/$\pi$ interactions
between the $\pi$ orbitals of the benzene rings of the plumbagin and
the $\sigma$ orbitals of the Hydrogen atom of the BCDs.
Although the structures in Figure \ref{fig.strComp} are relaxed
by CAM-B3LYP-D3, the functionals other than (CAM-)B3LYP predicted
the same patterns for the six conformers.
This figure shows that the BCD's conformers are the different patterns
each other, while those of MBCD [2-O-HPBCD] are the same pattern (a) [b].
Therefore, in order to get a reliable estimation of the relative energy
of the BCD's conformers, the functional has to be able to accurately
describe the totally different interactions.
We consider this difficulty is the reason why the functional dependent
contradiction just appears in the case of BCD.
This could suggest that, although the D3 correction and long-ranged
correction can comparatively accurately predict the binding energy,
further improvement on the truly functional part is needed to describe
the energy difference of the largely different molecular configurations.

\section{Conclusion}
\label{sec.conc}
We evaluated the reliability of the several functionals 
for the binding energies between cyclodextrins (BCD, MBCD, and 2-O-HPBCD)
and plumbagin. Comparing the functionals without D3 correction,
we established that Minnesota functionals qualitatively reproduced 
the stabilization by the binding, while (CAM)-B3LYP do not. 
This could be because the latter functionals are trained just
for covalent systems.
In our all tested functionals, B3LYP-D3, CAM-B3LYP-D3, M06-2X-D3,
and M06L quantitatively reproduced our DMC result.
Yet we concluded that the success of M06L is merely a fortune
due to the overbind by the unsatisfactory consideration of
long-range exchange interaction and the underbind by the lack of
the dispersion force correction cancelling out each other.
Focusing on the relative energy prediction between two types of
conformers, I and II, we found that the functional gave contradictory
predictions just in the case of BCD.
We concluded that the contradiction comes from that the two types of
conformers of BCD are combined by totally different interactions; 
the functional has to be eligible for fulfilling the difficulty
that it can accurately describe those different situations.
Since we showed that the degree of freedom of the D3 parameters
would hardly affect the binding energy prediction using the bootstrap
analysis, the contradiction for the BCD case suggests that
the truly functional part has to be further improved to
accurately describe the molecular encapsulation process from DFT.

\section{Acknowledgments}
This research used resources of the Argonne Leadership
Computing Facility, which is a DOE Office of Science User
Facility supported under Contract No.DE-AC02-06CH11357,
and the Research Center for Advanced Computing Infrastructure
(RCACI) at JAIST.
We thanks for Dr. Anouar Benali for letting us use the former
calculation resource. 
T.I. is grateful for financial suport from Grant-in-Aid
for JSPS Research Fellow (18J12653).
K.H. is grateful for financial support 
from KAKENHI grants (17K17762 and 19K05029), 
a Grant-in-Aid for Scientific Research 
on Innovative Areas (16H06439 and 19H05169),
and PRESTO (JPMJPR16NA) and the Materials research
by Information Integration Initiative (MI$^2$I) project of 
the Support Program for Starting Up Innovation Hub
from Japan Science and Technology Agency (JST).
R.M. is grateful for financial supports from MEXT-KAKENHI (17H05478 and 16KK0097), 
from Toyota Motor Corporation, from I-O DATA Foundation, 
and from the Air Force Office of Scientific Research (AFOSR-AOARD/FA2386-17-1-4049).
R.M. and K.H. are also grateful to financial supports from MEXT-FLAGSHIP2020 (hp170269, hp170220).
\bibliographystyle{apsrev4-1}
\bibliography{references}
\end{document}